\documentclass[a4paper,11pt]{article}
\pdfoutput=1

\usepackage{jheppub} 

\usepackage[T1]{fontenc} % if needed

\usepackage{graphics}
\usepackage{graphbox,graphicx}
\usepackage{xcolor}
\usepackage{amsmath}
\usepackage{amssymb}

\newcommand{\cenuns}{CE$\nu$NS}
\newcommand{\ganess}{GanESS}

\newcommand{\fe}{$^{55}$Fe} 
\newcommand{\kr}{$^{83m}$Kr}

\newcommand{\mum}{$\mu$m}
\newcommand{\mus}{$\mu$s}
\newcommand{\gtwo}{$g_2$}
\newcommand{\mycomment}[1]{}

\title{\boldmath The Gaseous Prototype (GaP): a \ganess\ demonstrator.}

\author[a,b,1]{L. Larizgoitia,\note{Corresponding author.}}
\author[a, c, 1]{A. Sim\'on,}
\author[a]{E. Oblak,}
\author[a]{C. Echeverria,}
\author[a]{P. Dietz,}
\author[a]{A. Castillo,}
\author[a]{L. Donneger,}
\author[a, d]{Juan Jos\'e G\'omez-Cadenas,}
\author[a, d]{F. Monrabal}

\affiliation[a]{Donostia International Physics Center, \\ Paseo Manuel Lardizabal 4, 20018, Donostia-San Sebasti\'an, Spain}
\affiliation[b]{Department of Physics, \\ University of the Basque Country UPV/EHU, PO Box 644, Bilbao, E-48080, Spain}
\affiliation[c]{Instituto de F\'isica Corpuscular, CSIC \& Universitat de Val\`encia, \\ Calle Catedr\'atico Jos\'e Beltr\'an 2, 46980, Paterna, Spain}
\affiliation[d]{Ikerbasque, Basque Foundation for Science, \\ Plaza Euskadi 5, 48013, Bilbao, Spain}

\emailAdd{leire.larizgoitia@dipc.org}
\emailAdd{ander.simon@ific.uv.es}

\date{Received: date / Revised version: date}

\abstract{
The GanESS experiment will exploit the high-pressure noble gas time projection chamber technology to detect coherent elastic neutrino-nucleus scattering (\cenuns) at the European Spallation Source (ESS). The detector, able to operate at pressures up to 50 bar with different noble gases (Xe, Ar and Kr), will employ electroluminescence to amplify the ionization signal with the objective of reaching a threshold as low as 1-2 e$^-$, equivalent to $<$ 100 eV$_{\text{ee}}$. \\
The Gaseous Prototype (GaP) has been built to characterize the technique at the few-keV energy regime and to understand various aspects related to the technology. Concretely, it will be used to measure the quenching factor of the different mediums as well as to characterize the electroluminescence yield and detection threshold under different operational conditions. The present paper describes the Gaseous Prototype and its first results operating with gaseous argon at moderate pressures (up to 10 bar). A potential detection threshold lower than 2.9 keV has been observed following operation with a $^{55}$Fe calibration source.
} %end of abstract

\usepackage{lineno}
\begin{document}
%\setpagewiselinenumbers
%\linenumbers

\maketitle
\flushbottom

\section{Introduction}
\label{intro}

Predicted in the 1970s \cite{Freedman:1973yd}, coherent elastic neutrino-nucleus scattering (\cenuns) was only measured for the first time a few years ago \cite{COHERENT:2017ipa, Scholz:2017ldm}
In this process low energy neutrinos (few tens of MeV) interact coherently with an atomic nucleus as a whole, through the weak current channel, as long as the coherent condition |Q| $<$ 1/R is satisfied, where |Q| is the momentum transfer and R is the nucleus' radius. Due to the low value of the weak mixing angle for protons, the coupling ends up being effectively proportional to the squared number of neutrons (N$^2$) in the target nucleus. The observation was received enthusiastically as it opened the path to many phenomenological proposals based on the process which would strongly benefit from improved measurement. Still, current measurements are heavily limited by signal statistics and further developments are needed to 
thoroughly explore all the possibilities derived from \cenuns\ measurements.

The development of new neutron spallation sources, such as the European Spallation Source (ESS), and upgrades of existing ones provide an opportunity in the next years to fully exploit the physics of the \cenuns\ process with new technologies that could guarantee measurements not limited by statistics. In \cite{Baxter:2019mcx, Abele:2022iml} we already explored the capability of the ESS as a source to 
entirely exploit this process and its related physics.

The Gaseous Detector for Neutrino Physics at the ESS (\ganess) experiment will develop a high-pressure noble gas time projection chamber with electroluminescence amplification (HPNG EL-TPC) to observe the \cenuns\ process from spallation neutrinos (pion decay-at-rest process). The \ganess\ detector aims to observe for the first time the \cenuns\ process from spallation neutrinos in xenon and operate the detector also with different noble gases to allow full exploitation of measurements not limited by statistics, with the same systematics but different nuclei. The technique offers several attractive traits. First, it can be operated with different noble gases without any major intervention to the system and thus yields to \cenuns\ measurements in different targets with the same detector, xenon and argon being the main candidates for its operation. This will be greatly beneficial to constrain the parameter space for different physics scenarios \cite{Baxter:2019mcx}. Second, optical amplification of the ionization signal, via electroluminescence (EL), enables a potential detection threshold as low as the energy required to form an electron-ion pair (for reference, 22.1 eV in gaseous Xe \cite{RevModPhys.52.121}). Third and finally, while noble gas dual-phase detectors are affected by charge trapping in the inter-phase between liquid and gas \cite{RED-100:2019rpf, LUX:2020vbj, Baur:2022sel}, that's not the case for single-phase detectors. It should be noted that operation in the gaseous phase implies a considerably lower interaction rate when compared to other \cenuns\ detectors. However, this circumstance is mitigated thanks to the large neutrino flux at the ESS and the operation at high pressure.

\begin{figure}[!h]
\centering
  \includegraphics[width=0.6\textwidth]{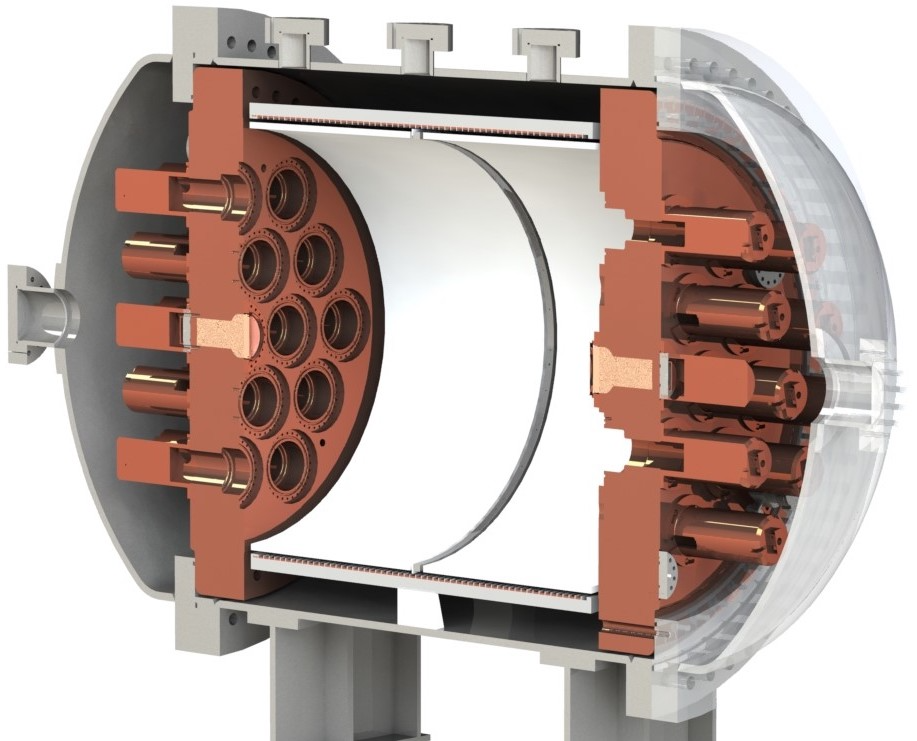}
\caption{Conceptual design of the GanESS detector. It consists of two symmetric TPCs with a central solid cathode with 60 cm diameter and 30 cm drift length. Two planes of PMTs will detect with high efficiency the light produced in the EL regions right in front of them.}
\label{fig:ganess}
\end{figure}

The baseline design of \ganess\ is shown in Fig.~\ref{fig:ganess}. It will be a cylindrical symmetric TPC with two 30 cm long drift volumes separated by a central cathode and a diameter of 60 cm. This geometry will hold $\sim$20 kg of Xe when operating at 20 bar. With this mass, when the ESS operates at its full power, $\sim$7,000 \cenuns\ Xe events per year will be detected at a distance of 20 m from the ESS target with a threshold of 1 $\rm{keV}_{\rm{nr}}$ assuming a 20\% quenching factor \cite{Baxter:2019mcx}. The number is reduced to $\sim$700 events per year when operating with Ar at the same pressure. %\todo{redo numbers with 24m? Ask Leire}.

In this paper, we describe in detail for the first time the Gaseous Prototype (GaP) detector at DIPC, section \ref{sec:gap}. Data acquisition and processing are discussed in section \ref{sec:daq}, followed by results in section \ref{sec:results}.

\section{The Gaseous Prototype}\label{sec:gap}

Broadly used for neutrinoless double beta decay searches in the context of the NEXT experiment \cite{NEXT:2018rgj, NEXT:2015wlq}, the HPNG EL-TPC technique has been primarily developed for moderate pressures, up to 15 bar; lower than \ganess\ goals. Moreover, existing similar detectors have been optimized for signals in the few MeV range, which is higher than the energy region of interest for \cenuns\ signal (sub to few keV). The performance of noble gas TPCs in such regime is yet to be explored and needs to be characterized to fully assess and understand the potential of the technique for low-energy searches.

\begin{figure}[!tbh]
\centering
\resizebox{0.75\linewidth}{!}{%
  \includegraphics{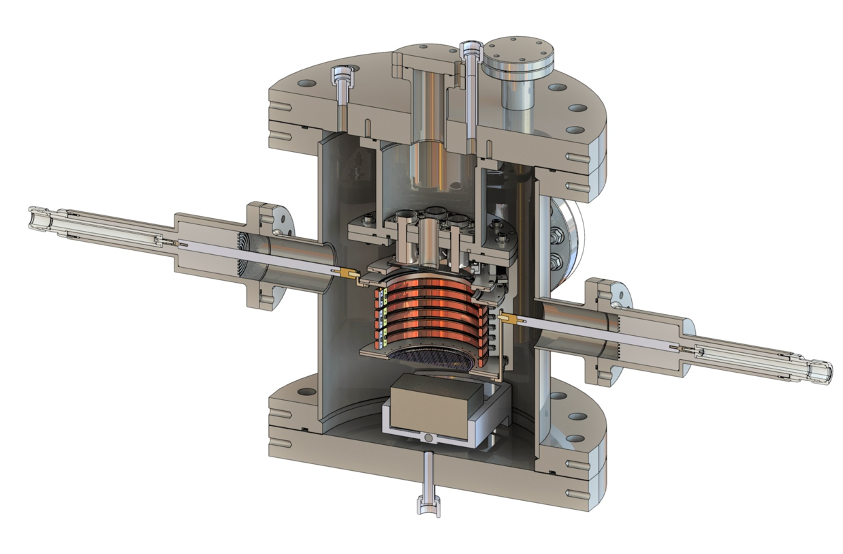}
}
\caption{Render image of the inner parts of the GaP detector. The field cage and high-voltage feedthroughs are clearly visible with the EL region on the upper side of the detector. Just in front of the EL there is a plane of seven 1" PMTs. While the detector is designed to allow for pressure isolation of the PMTs, in this initial run they were in contact with the gas and then the operation has been limited to pressures below 10 bar.}
\label{fig:gap1}
\end{figure}

The Gaseous Prototype (GaP), a small time projection chamber built at the Donostia International Physics Center in 2023, aims to understand many of these aspects. The detector, described in the following and illustrated in fig.\ref{fig:gap1}, has been operated over the past months with gaseous Ar at various pressures up to 10 bar.

\subsection{Time projection chamber}\label{gap:tpc}

\begin{figure}[!tbp]
\centering
\includegraphics[height=5.5cm]{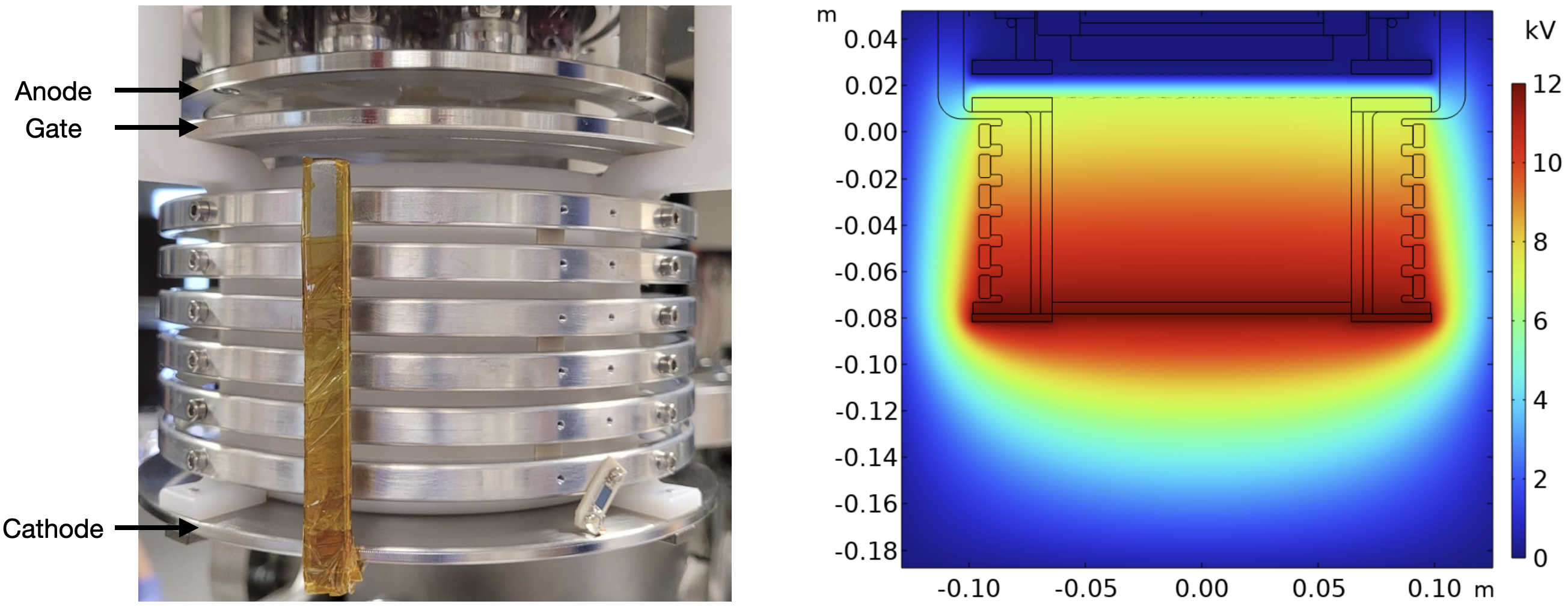}
\caption{Left: Full view of the GaP TPC. The cathode, gate and anode are visible while 6 rings connected by resistors guarantee the creation of a uniform electric field in the central part of the detector. The metallic piece wrapped with kapton allows to connect the High-Voltage feedthrough to the cathode.
Right: Comsol simulation of the electric field inside the GaP vessel. The simulation shows the homogeneity of the electric field in the active region of the detector generated by the field cage shown in this figure.}
\label{fig:gap_TPC}
\end{figure}

The time projection chamber, shown on fig.~\ref{fig:gap_TPC} left, is composed by three stainless steel electrodes: a cathode, a gate and a grounded anode. The cathode is a solid plate with mechanized apertures to facilitate the recirculation of the gas within the chamber volume. It also has a central hole used to place radioactive sources, which are held in place by an additional plate. The gate is a thin wire mesh of 50 \mum\ diameter wires with a 500 \mum\ spacing (0.81 transparency) that has been cryo-fitted to two concentrical rings acting as holders. The anode is a 75 \mum\ thick photoetched hexagonal grid attached to a holder ring with kapton tape in the edges. Its hexagons have a 1.467 mm side and their contour is 150 \mum\ wide resulting in a 0.89 transparency.

The three electrodes divide the chamber into two areas: the drift and the electroluminescence regions. The drift volume is limited by the cathode and gate, which are separated by 8.7 cm. A moderate voltage difference between them results in an electric field that guides the electrons toward the electroluminescence (EL) region, avoiding electron recombination following an ionization event. To ensure uniformity, the cathode and gate are electrically connected via a resistor chain of 100 MOhms resistances. Starting from the cathode and ending in the gate, each resistance is connected in series to six field-shaping aluminum rings of 130 mm inner diameter, 5 mm thickness and 10 mm width. The electroluminescence region, limited by the gate and the anode, spans over a nominal separation of 1.02 cm. There, a much higher electric field is applied, which accelerates the electrons enough to excite the gas and produce secondary scintillation via electroluminescence. Both the drift and electroluminescence field have been simulated with COMSOL Multiphysics\textsuperscript{\textcopyright} finite element software and are shown in Fig.~\ref{fig:gap_TPC} right.

Two Fug HCP35-35000 High-Voltage power supply modules are used to apply voltage to the cathode and gate. These modules are controlled and monitored with a custom slow control system, built in LabView \textsuperscript{\textcopyright}, which allows to increase the voltage gradually in small steps to avoid discharges in the ramp up. To protect the system, the software turns off the voltages automatically in the event of a discharge.

Two identical high voltage feedthroughs (HVFTs) are used to connect the power supplies to the cathode and gate. Given the short drift length, the difference in voltage between the cathode and gate is marginal compared to the electroluminescence requirements; therefore, the voltage requirements are defined by the electroluminescence threshold. As the goal of GaP is to operate up to 50 bar, they must be able to hold up voltages of at least $\sim$42 kV, slightly above the 0.83 kV/cm electroluminescence threshold of Xe \cite{Monteiro:2007vz}. The design of the feedthroughs is based on the one used for NEXT-White's gate \cite{NEXT:2018rgj}, which was able to hold up to 22 kV at 15 bar. The design relies on the gas itself to provide insulation and employs a metal rod with a spring contact to the electrodes. The connection to the gate is done directly with the feedthrough. However, that is not the case for the cathode, as its position does not coincide with the port position. A metallic l-shaped piece, screwed to the cathode, makes the connection between the cathode and the spring contact. Outside the contact regions, the piece is wrapped in kapton to avoid electrical discharges with the field rings. The anode is electrically connected by 4 metallic pieces, which also act as holders, to the rest of the pressure vessel, which is grounded. The rest of the TPC is mounted vertically on 4 high-density polyethylene rods which are used to attach the TPC to the pressure vessel without making electrical contact.

A hollow cylindrical PTFE tube is inserted inside the drift volume to maximize light collection (light tube). The top side, facing the EL region, is open while the bottom one is solid with the same apertures as the cathode, where it sits. The thickness of the cylinder is 5 mm and its inner radius is 60 mm. The tube is coated with tetraphenyl butadiene (TPB), a wavelength shifter commonly used to shift the VUV light from noble gas scintillation towards blue ($\sim$425 nm) in order to maximize reflectivity and light collection efficiency \cite{MCKINSEY1997351}. The coating was done in-house with a tabletop evaporator. Although the system lacks precise control of the coating thickness, a few micrometer thickness coating is expected, sufficient for efficient shifting \cite{Benson:2017vbw}. 

\subsection{Gas system and pressure vessel}\label{gap:gas}

\begin{figure*}
\centering
  \includegraphics[height=7.5cm]{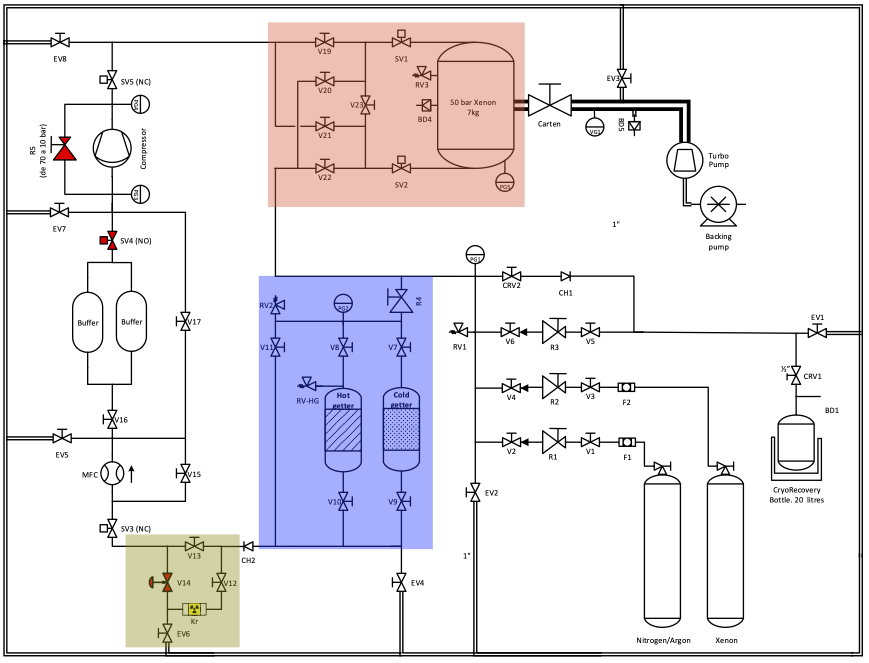}
  \hspace{0.7cm}
  \includegraphics[height=7.5cm]{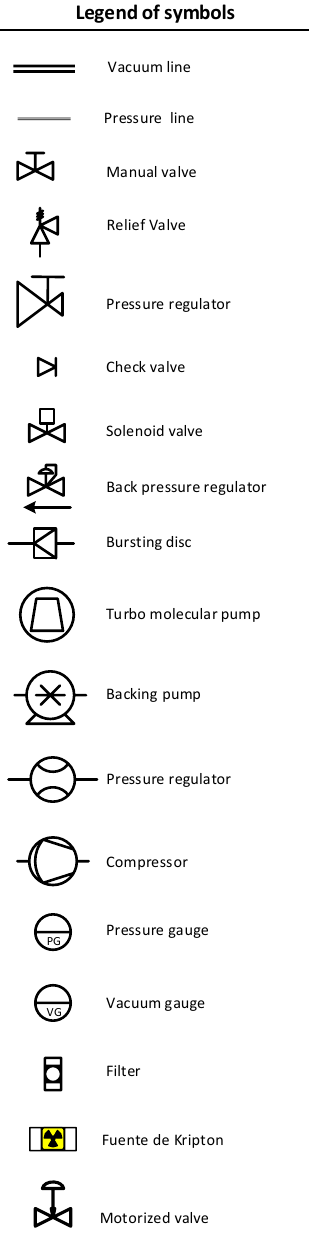}
\caption{GaP gas system scheme. The red area covers the GaP vessel and the valve system that allow to change the gas circulating direction in the detector. The blue area covers the getters section of the gas system. As this section cannot operate at pressures above 10 bar it is equipped with a regulator at the entrance of it and with a non-return valve at the exit allowing for a separate pressure section. The green area represents a part of the gas system where a gaseous Rb/Kr source can be installed.}
\label{fig:gas_system}
\end{figure*}

Operation with clean gas is mandatory in noble gas TPC as impurities reduce the electron lifetime and quench VUV light. To maximize gas purity, the pressure vessel, which houses the TPC, is connected to a recirculation loop, detailed in Fig.~\ref{fig:gas_system}, which includes gas purifiers. During operation, gas is in continuous recirculation through this loop. Recirculation is forced by one of two different pumps, each used to operate in a different pressure regime. For low pressure, up to 10 bar, a smaller single diaphragm pump is used. At higher pressures, recirculation is produced by a 2-stage compressor by SERA company, which operates to a maximum of 50 bar. As the purifiers cannot operate above 10 bar, the gas system is designed with a low-pressure region where the gas circulates below this pressure and is purified. The gas is then pumped up by the SERA compressor and re-introduced into the detector. In order to prevent large decreases in pressure at the entrance of the compressor, two buffer volumes are installed at its entrance increasing the volume of the low-pressure region.
The system counts with a series of vacuum lines and connections that allow for an efficient evacuation of the whole volume prior to its filling with a noble gas using a combination of a scroll and turbo pumps. After air evacuation, gas is introduced into the system via 3 different inlets, each connected to a different bottle, facilitating gas changes without major interventions. An outlet for cryo-recovery is also available and it will be used when operating with pure xenon.

An ambient getter (Sigma Technologies MC1500-902 FV model) and a hot getter (Sigma Technologies PS4 MT15 R2 model) are placed in parallel in the recirculation loop, allowing operation with either one or both of them at the same time. They are placed before the inlet of the recirculation pump as they are not able to operate at the maximum operational pressures of GaP. Before them, a pressure regulator reduces the pressure to a maximum of 10 bar. This regulator, combined with a non-return valve at the exit of the getters region, allows to protect this part of the gas system and ensures the safe operation of the getters when operating at higher pressure.
The gas system also includes a section that allows to introduce a \kr\ source, used for detector characterization \cite{NEXT:2018sqd}. A general vacuum line is connected to the gas system at various points. This configuration, in combination with the different valves distributed throughout the system, allows individual pumps in different sections if required. Pressure and vacuum are monitored at different points of the system using a set of gauges read by a custom slow control module developed on LabView\textsuperscript{\textcopyright}.

\begin{figure}
\centering 
\includegraphics[width=0.75\linewidth]{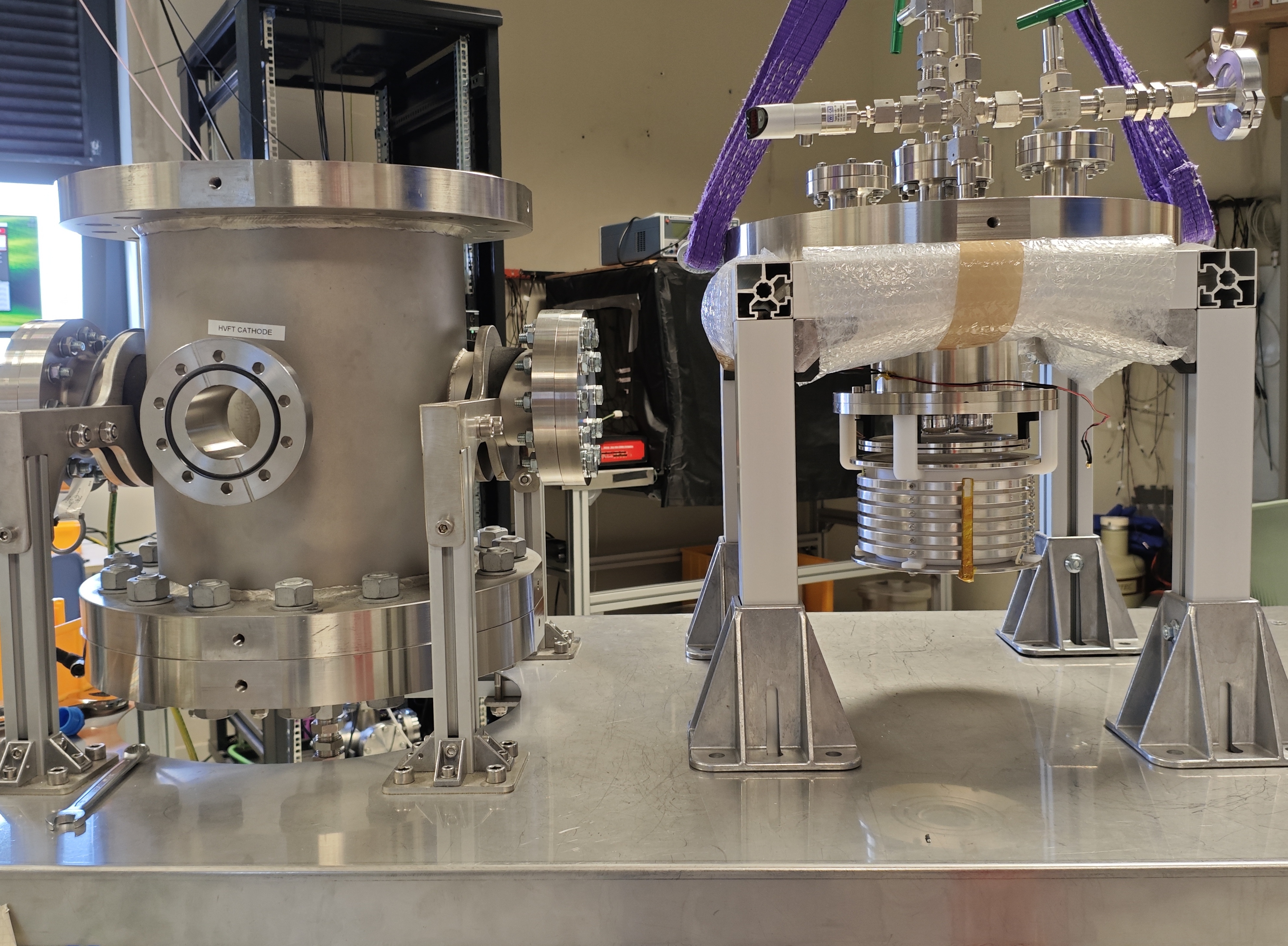} \hspace{1cm}
\caption{GaP pressure vessel and inner detector. The lateral ports are used for the HVFTs. The top endcap, attached to the field cage in the image, the central port is used to power the PMTs and extract their signal while the lateral ones are used to pulse an internal LED for calibration. The field cage is shown for comparison with the vessel size.}
\label{fig:vessel} 
\end{figure}

The TPC is housed inside a 6-mm-thick cylindrical stainless steel pressure vessel of 13.8 cm internal radius and 38.65 cm length, capable of holding up to 50 bar, Fig.~\ref{fig:vessel}. An additional cylindrical inner chamber is fixed to the top cap of the vessel. This volume has seven holes in the bottom to accommodate an equal number of photomultiplier tubes (PMTs) for light detection. Although not currently installed, a quartz window can be coupled to isolate the sensors, rated for up to 20 bar, from higher pressures. In addition, the TPC hangs from this inner cylinder.

The vessel is connected to the recirculation system through two gas lines, one located at the bottom and another at the top. The recirculation direction can be changed by manipulating a system of valves. The top connection is done through a cross, which is also connected to a pressure gauge to monitor the pressure vessel, and a KF port connected to a vacuum pump to clean the vessel prior to filling the detector. An additional gas line is connected to the volume of the inner cylinder and it is used to monitor the vacuum inside its volume.

The vessel has 4 lateral ports distributed uniformly, i.e. each port is at $\pm$ 90 degrees from its neighbors. Two of the ports, located each in front of the other, are used for the high-voltage feedthroughs while the other two are left available for future use. Three additional ports are in the top cap.  One of them is centered, connected to the inner cylinder, and it is used as a feedthrough for the photosensor supply and signal. The other two are placed at a larger radius, communicating directly to the main volume. Thick methacrylate windows with a 375 nm LED coupled to them were used initially at these ports during sensor calibration periods to prevent noise coupling to the PMTs. This was solved in a later operation of the detector by using an extra feedthrough for the LED.

\subsection{Photosensors and DAQ system}\label{gap:sensors}

\begin{figure}
\centering
\resizebox{1\linewidth}{!}{%
  \includegraphics{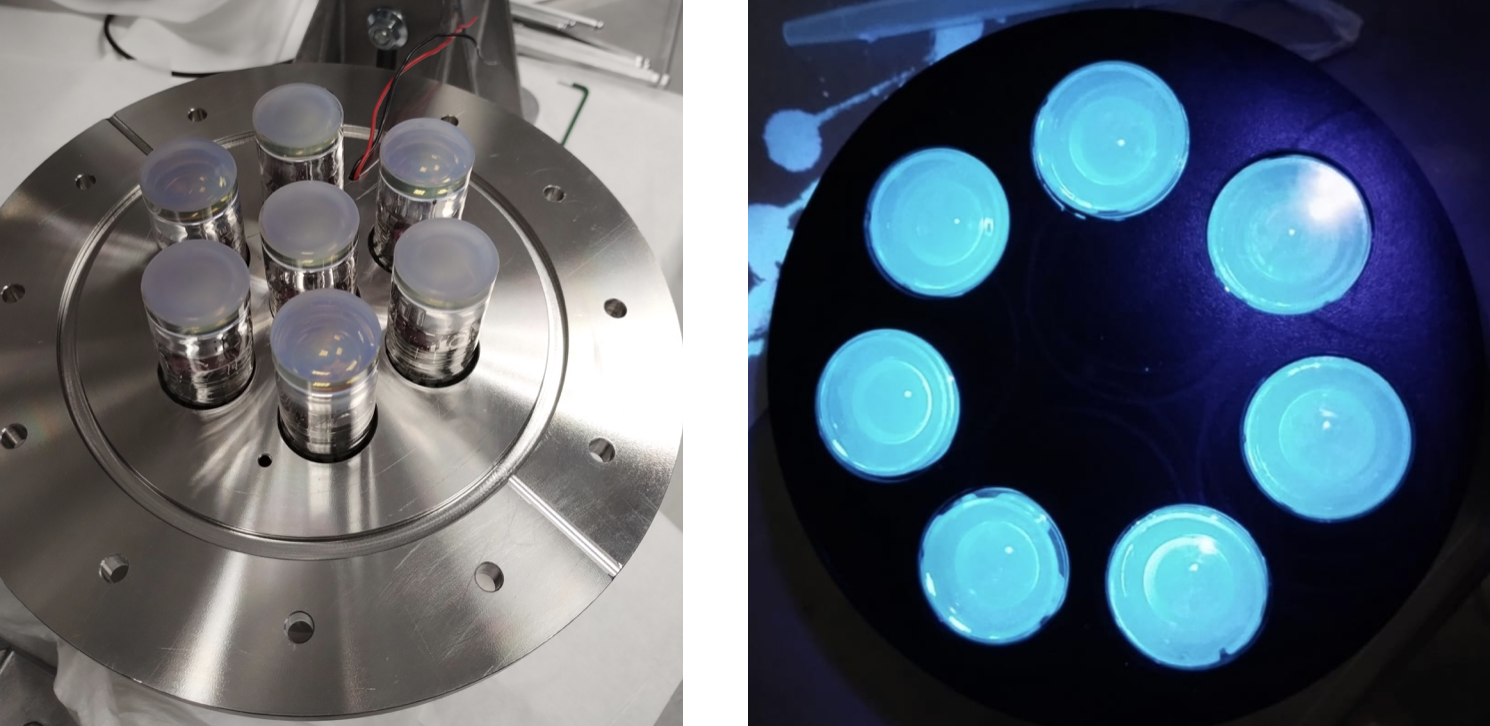}
}
\caption{Left: GaP PMTs distribution. Right: PMTs illuminated with UV light after deposition of TPB.}
\label{fig:gap_pmts} 
\end{figure}

Light is detected by seven 1-inch head-on Hamamatsu R7378A photomultiplier tubes located behind the anode grid. The sensors are distributed hexagonally, with one of them placed in the center of the hexagon and the rest at 36.37 mm from the center as illustrated in Fig.~\ref{fig:gap_pmts} left. The distance between the sensors and the grid can be adjusted thanks to the space available inside the inner cylinder, where the PMT bases are fixed. For the current configuration, where their maximum rated pressure (20 bar) was not exceeded, the PMTs were partially placed inside the inner cylinder, with their face 2 mm away from the anode grid to maximize the solid angle covered by the sensors. 

The quantum efficiency of the sensors at the Xe scintillation wavelength (175 nm) is rather low, $\sim$10\%, and they are directly non-sensitive to Ar light. For this reason, we evaporated TPB directly into the PMT window. Fig.~\ref{fig:gap_pmts} right shows the TPB coated PMTs illuminated with a UV lamp. On the other hand, the light emitted towards the cathode will be shifted on the PTFE reflector walls, also coated with TPB.

The PMT bases design follows the manufacturer's recommendations, with several resistors, capacitors and pin receptacles mounted on FR-4 boards. The bases are fully covered in epoxy to avoid dielectric breakdown at low pressures. Four kapton wires, two for biasing and two to extract the signal, are connected to each base. These are connected to a 36-pin feedthrough attached to the central top port. A 375 nm LED, intended for sensor calibration, is linked to a lateral feedthrough. The initial operation was performed using a feedthrough rated to pressures up to 10 bar, which currently limits operational pressure. A new design for this feedthrough is already tested and available to allow operations up to 50 bar for the second run. The PMTs are negatively biased with a CAEN A7236SN power supply and operated at 1250 V, for an average gain of 10$^6$. 

Signals are directly read with a CAEN DT2740 digitizer. It is a 64-channel tabletop unit with a 16-bit 125 MS/s ADC, i.e. 8 ns time binning, and an input range of 2 V$_{pp}$. While the system allows for custom programming of the FPGA, the WaveDump2 \cite{wavedump} software was used to acquire data. It allows for simultaneous acquisition of all active channels following an edge trigger signal whose level can be independently configured for each channel. A maximum pretrigger of 16 \mus, corresponding to 2000 samples, is allowed by the software, while the buffer size can be expanded over several hundreds of \mus. From each channel, individual waveforms are acquired in ASCii format and later converted to \textit{hdf5} format to reduce their size.

\section{Data taking and processing}
\label{sec:daq}

Since being mounted, GaP has been operated with Ar at pressures ranging roughly from 1.5 to slightly above 8.5 bar. Different calibration sources have been used to characterize the detector response. Moreover, data was taken with and without the light tube.

While specific details will be given in the following, generally speaking, data was taken in short periods, up to 15 minutes, for each given configuration to avoid time-dependent variations. Each configuration had its own set of operational parameters which included a fixed pressure, cathode and gate voltages. Gas was continuously recirculated during data-taking and initiated daily at least one hour prior to the start of data-taking in order to remove impurities outgassed during the night when the recirculation loop was stopped to prevent possible unsupervised damages to the system. The estimated time for the gas to complete a whole loop through the system is about 15 min, with one hour of circulating before the initial data taking the gas proven to be clean enough and in stable conditions.

\subsection{\fe\ internal source}
\label{subsec:Fe}
\begin{figure}
\centering
\includegraphics [trim={0.2cm 0 0 0},clip]
{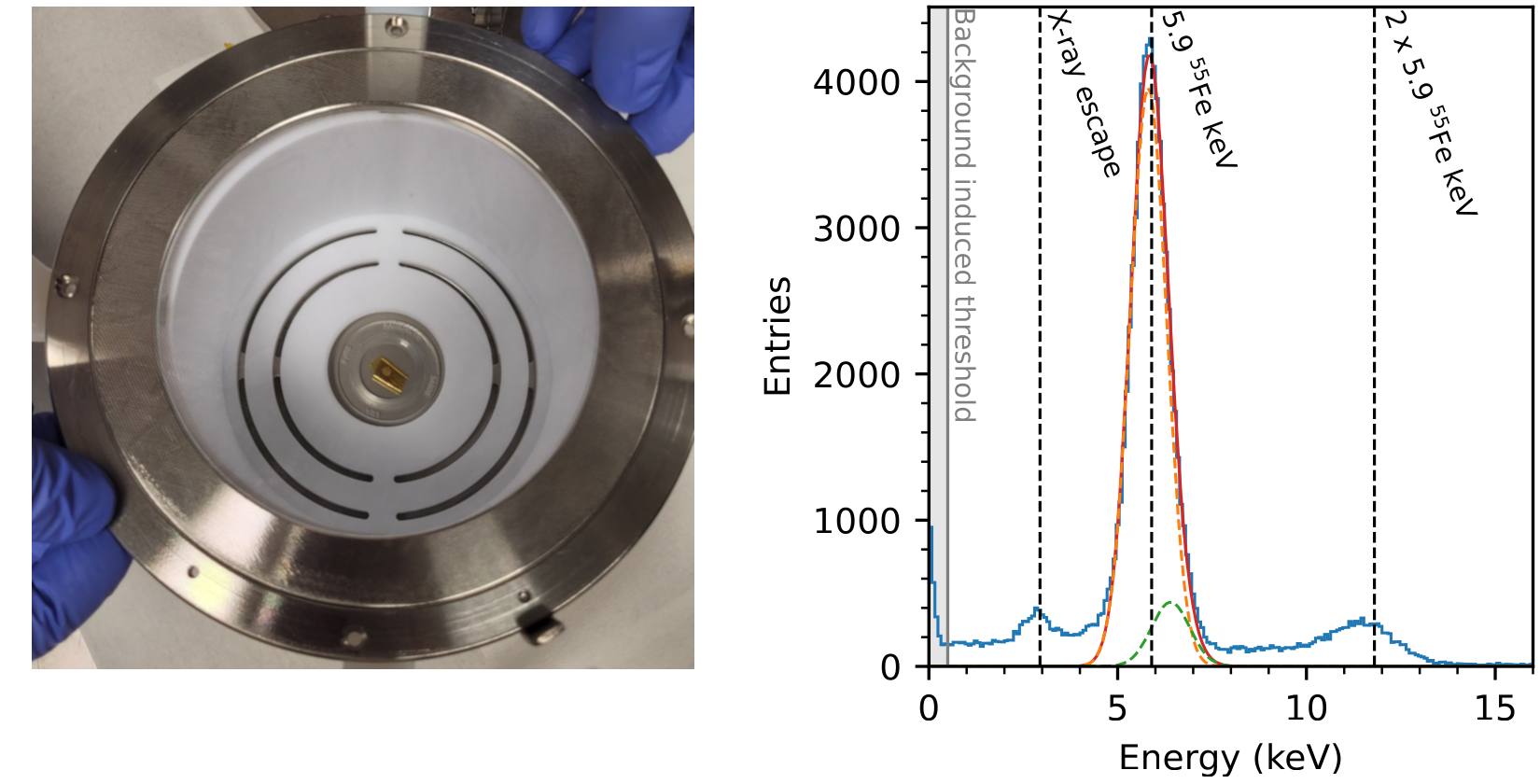}
\caption{Left: Image of the position of the \fe\ source in the detector. The source is attached to the cathode with a small kapton tape. The position corresponds to the radial center of the active volume. Right: Energy spectrum obtained from the \fe\ source. A clear peak, corresponding to an energy of 5.9 keV, the dominating line, is seen. The red line corresponds to the result of the two Gaussian fit described in the text, while the dashed lines correspond to the 5.9 (orange) and 6.5 (green) keV populations. A duplicate of the peak is seen at twice the energy, corresponding to events which had accidental coincidences due to the high activity of the source. At 2.94 keV a peak corresponding to the escape of Ar $K_{alpha}$ X-rays are also seen. In gray, a low-energy background population that appears at high electric fields and limits the detection threshold to $\sim$0.5 keV. The detection threshold will be discussed in \ref{subsec:det_threshold}.}
\label{fig:fe_spectrum} 
\end{figure}

A \fe\ source was placed in the cathode's central hole. \fe\ decays into $^{55}$Mn via electron capture. The decay is followed up by the emission of Mn characteristic X-rays, with energies 5.9 and 6.5 keV corresponding to the $K_{\alpha}$ and $K_{\beta}$ lines.
The source, provided by ORANO, is a 3 mm diameter radionuclide disk placed between 2 thin polyester foils of 75 \mum\ thickness, mounted on a 38 mm diameter plexiglass ring (Fig. \ref{fig:fe_spectrum}). The original activity of the source was 20 kBq when bought (June 2023) and is estimated to be 15.85 kBq at the moment of data-taking (May 2024). 

Data was acquired by triggering on the central PMT as we expect it to be the sensor with a larger signal, being closer to the radial position of the source. The trigger threshold was set to a fixed value slightly above the electronic noise. Given the low energy of the gammas, the trigger was only sensitive to secondary scintillation. 

The mean free path of the gammas will range between 0.24 and 1.1 cm, depending on the operational pressure. This means that their interaction point will be extremely concentrated around the source.
However, the distance between the interaction point and amplification region precludes the observation of primary scintillation. The drift velocity for Ar at these pressure ranges was calculated using Magboltz \cite{Biagi:1999nwa}, the calculation indicates a drift velocity between 1 and 3 mm/\mus. At these conditions, the primary scintillation will reach the photosensors several tens of \mus\ before the start of the secondary scintillation. As the maximum pre-trigger of the system is 16 \mus, the primary scintillation cannot be observed when triggering on S2. With this in mind and to maximize the acquisition rate, only limited by the data throughput, the acquisition window was set to 40 \mus\ with a pretrigger of 10 \mus.

\begin{figure}
\centering
\includegraphics{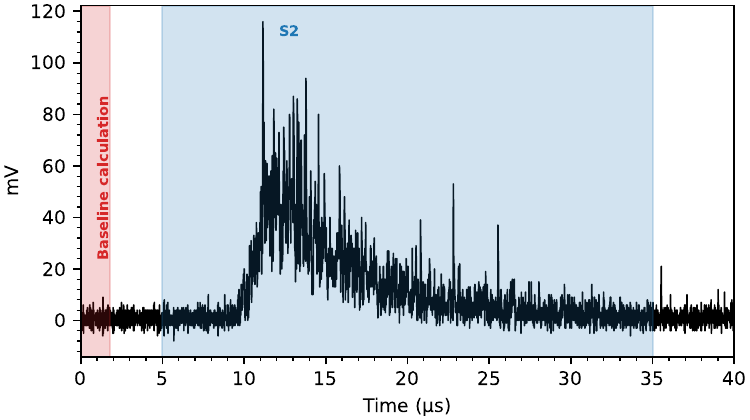}
\caption{A typical waveform signal produced by the secondary scintillation of \fe\ induced interactions. Data is triggered at 10 \mus, close to the start of the pulse. The red region covers the 225 samples used to calculate the baseline. The integration window, shaded in blue, expands from 5 to 35 \mus. This interval contains more than 98\% of the integrated charge of the full waveform while the losses are related to delayed emissions, like the small peak at 35.5 \mus.}
\label{fig:fe_wvf} 
\end{figure}

A simple data processing scheme is applied to create analysis datasets. For each event, the baseline of each channel is calculated, as the average of the first 225 samples (first 1.8 \mus ), and is later subtracted from the signal. The signal of each channel is later converted to photoelectrons based on calibration factors periodically obtained with an LED. It should be noted that a constant decrease in such conversion factors was observed over several months. This variation is considered a systematic uncertainty in the analysis. 

Following conversion into photoelectrons, the waveforms of the 7 PMTs are added together. The integral of such sum in the interval [5, 35] \mus is taken as the recorded energy of the event. The considered interval is enough to cover the signal, as shown in Fig.~\ref{fig:fe_wvf}. The resulting spectrum obtained is shown in Fig.~\ref{fig:fe_spectrum}. A fit to two Gaussians, corresponding to the peaks at 5.9 and 6.5 keV, is applied. The centroid of the higher energy peak is constrained to be a factor 6.5/5.9 larger than the lower energy peak. The centroid value of the lower energy peak is considered as the average number of detected photons, $N_{det}$, for 5.9 keV depositions, which can be used, as discussed later, to evaluate the charge yield of the detector. 

The energy resolution, obtained from the fit to the energy spectra at different pressures and electroluminescence fields, is shown in Fig.~\ref{fig:reso}. It shows a minimum value at a reduced electric field of 1.3-1.5 kV/cm/bar with a considerable worsening for low pressures mostly related to the lower number of photons produced in the amplification process. The resolution obtained is worse than anticipated by the number of photoelectrons detected per electron, this will be discussed in section \ref{subsec:eres}.

\section{Results and discussion}
\label{sec:results}

\subsection{Charge yield and light collection efficiency}
\label{subsec:charge_yield}

The charge yield \gtwo\ of the system is defined as the number of photoelectrons detected per ionization electron. It is one of the key aspects of the detector as it affects the energy resolution and the detection threshold being the threshold inversely proportional to \gtwo. We define \gtwo\ as:

\begin{equation}
    g_2 = \frac{N^{S2}_{det}}{N_{e^-}} = \frac{N^{S2}_{det}\cdot W_i}{E}
    \label{eq:g2}
\end{equation}

being $N^{S2}_{det}$ the number of detected photoelectrons from the secondary scintillation light production for depositions of a given energy $E$, $N_{e^-}$ the number of produced electrons by such energy deposition and $W_i$ the average energy to produce an ion-electron pair, also known as the $w$-value, which is taken as 26.27$\pm$0.14 eV \cite{wi_argon}. On the other hand, in our system, $N^{S2}_{det}$ is given by the product of the secondary scintillation light collection efficiency ($LCE_{S2}$, defined as the probability of detecting a photon produced via electroluminescence), the electroluminescence yield ($Y_{EL}$, number of photons produced per ionization electron) and the number of electrons arriving at the amplification region: 

\begin{equation}
    N^{S_2}_{det} = LCE_{S2}\cdot Y_{EL}\cdot N_{e^-}
    \label{eq:LCE}
\end{equation}

It follows that, ultimately, \gtwo\ is proportional to $LCE_{S2}$ and $Y_{EL}$. The light collection efficiency is a purely geometrical factor and should not, to first order, change with neither the gas pressure nor the electroluminescence voltage. On the other hand, the electroluminescence yield has been commonly described as a linear process that depends on the electric field applied in the amplification region. 

\begin{figure}
\centering
  \includegraphics{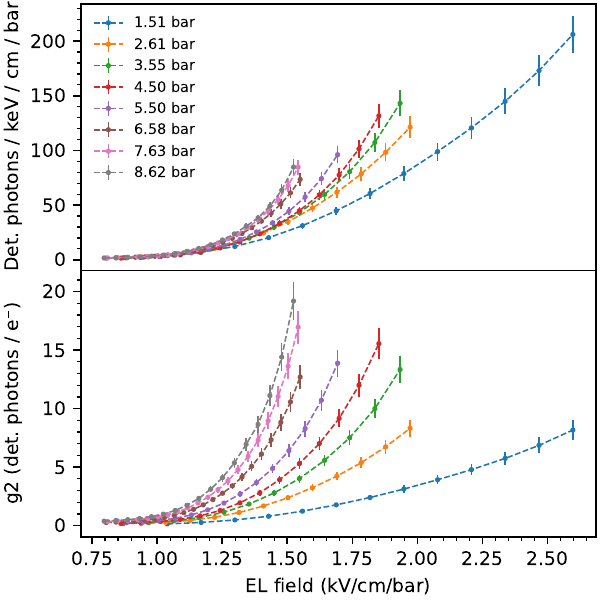}
\caption{Detected photons per keV (top) and \gtwo\ (bottom) as a function of the electroluminescence field for different operating pressures. An exponential trend is observed at all pressures. The difference in the number of det. photons per bar indicates that the reduced yield depends not only on the reduced field but also on the absolute pressure, indicating an extra light production mechanism in the EL region.}
\label{fig:charge_yield} 
\end{figure}

However, we observe an exponential increase of \gtwo\, illustrated in Fig.~\ref{fig:charge_yield}, as a function of the electroluminescence field. As explained above, this behavior is counter-intuitive as \gtwo\ should follow the same trend as the electroluminescence yield, which should behave linearly up to $\sim$3 kV/cm/bar. 

 An exponential behavior could be related to operation at sufficiently high electric fields to produce charge amplification, at least in some parts of the amplification structure like wires or the borders of the electroformed mesh. In an attempt to understand the trend, the electric field has been simulated with COMSOL\textsuperscript{\textcopyright} and used as an input for Garfield++ \cite{GarfieldPP} to evaluate the light yield. The details of this study are discussed in \ref{subsec:el_sim}.

\subsection{Electroluminescence simulation}
\label{subsec:el_sim}

As indicated in the previous section, in order to evaluate the effect of possible charge amplification in the surroundings of the EL wires due to a rapid increase in the electric field we have run microscopic simulations of the electron interaction with the gas using a combination of Garfield++ libraries with an electric field in the region calculated by finite element software COMSOL \cite{comsol2024}. 

\begin{figure}[tbp]
\centering
  \hspace{-0.1in}
\includegraphics{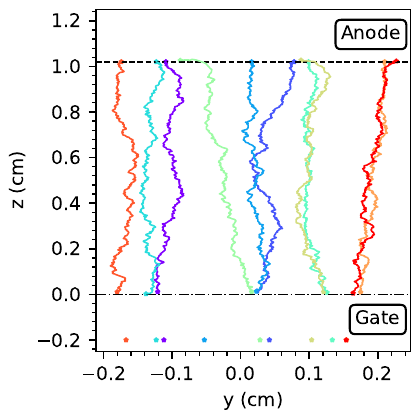}
  \hspace{0.1in}
  \includegraphics{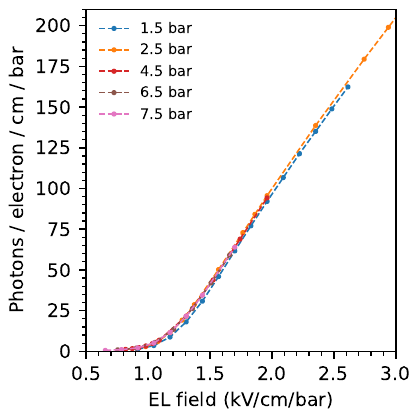}
\caption{Left: Illustration of the electron trajectories through the EL region simulated by Garfield++ for different starting points of the electrons. The Gate and Anode are represented by the bottom and top discontinuous lines respectively. The plot represents every interaction of the electron in the gas where it produces an excitation. The simulation is run for 1.7 kV/cm/bar at an absolute pressure of 7.5 bar. 
Right: Light yield in the EL as a function of the reduced electric field for different pressures according to Garfield++ simulation. We can observe the expected linear behavior at least up to 3 kV/cm/bar.}
\label{fig:garfield1} 
\end{figure}

The geometry simulated for the EL region corresponds to the characteristics of the meshes used in this data taking, a hexagon mesh in the anode and a woven mesh in the gate as described in section \ref{gap:tpc}. The mesh generated by COMSOL to perform the calculation had as the smallest size a fraction of the thickness of the hexagons to guarantee proper electric field calculations in these areas. We then run the simulations for different pressures and reduced electric fields to compare with the data at similar operation conditions of our detector. An example of different electrons propagating across the EL is shown in Fig.~\ref{fig:garfield1} left.

\begin{figure}[tbp]
\centering
  \hspace{-0.1in}
  \includegraphics{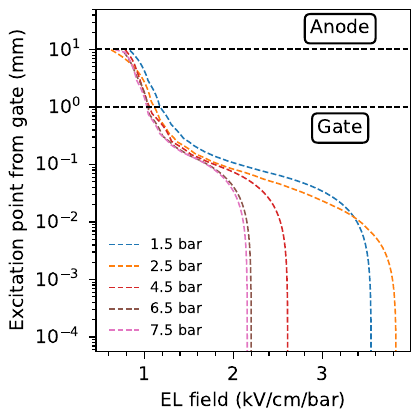}
  \hspace{0.1in}
  \includegraphics{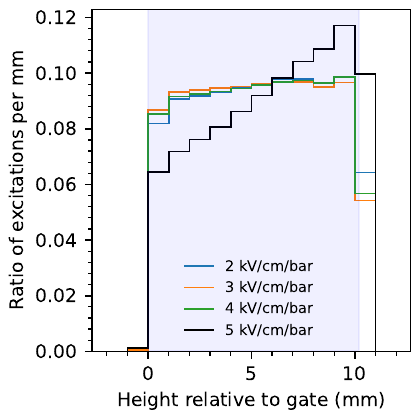}

\caption{Left: Mean value of distance from the gate for the interactions producing an excitation. The image shows that above $\sim$ 1 kV/cm/bar the first excitation occurs in the first millimeter of the EL. Above this value, the simulation predicts interactions every fraction of a millimeter. 
Right: Fraction of excitations produced per millimeter for different reduced electric fields. The amount of excitations is almost constant until very high fields where ionization effects appear and the amount of light produced is larger near the anode.}
\label{fig:z_excitation} 
\end{figure}

\begin{figure}
\centering
  \hspace{-0.3in}
\includegraphics{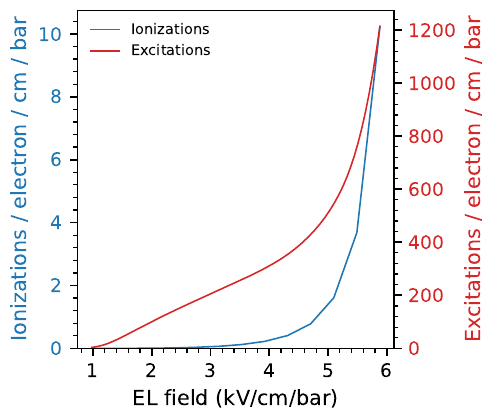}
\includegraphics{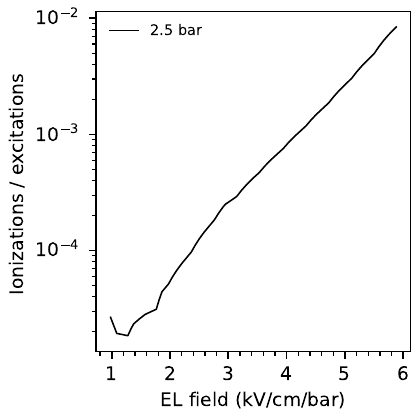}
\caption{Left: Number of predicted ionizations (blue) and excitations (orange) as a function of the reduced electric field. Up to the ionization threshold ($\sim$3 kV/cm/ba) only excitations occur. When ionizations are produced the reduced light yield increases exponentially as it is given in terms of primary electrons. Right: Fraction of the number of predicted ionizations over the number of excitations for different values of the reduced electric field. We can observe that this fraction, even while it increases rapidly with the field remains negligible (less than 10$^{-2}$ until very high fields, near 5 keV/cm/bar. Visible fluctuations in the low El field range were to be expected as it is below the ionization threshold. 
}
\label{fig:ioni_exc_fraction} 
\end{figure}

We then count the number of excitations produced per electron (corresponding to the light yield) and the number of ionizations to check for possible gain effects.
We also evaluate the z-position of the first excitation produced to look for possible optical differences in photon detection.
A summary of the results can be seen in Fig. \ref{fig:z_excitation} and \ref{fig:ioni_exc_fraction}.
From these results we can observe that the predicted number of photons produced is linear even at rather large reduced electric fields (Fig.~\ref{fig:garfield1} right), refusing the hypothesis of extra ionizations contributing to the total light in the regions near the wires.
Also, Fig. \ref{fig:z_excitation} shows that the location of the first excitation is similar for all configurations and the number of excitations produced is very similar along the EL region (fig. \ref{fig:z_excitation} right), discarding also any possible optical effects at larger electric fields.
Finally, the fraction of ionizations potentially contributing to light production remains very small for moderate electric fields, Fig. \ref{fig:ioni_exc_fraction} right. Indeed, the left figure shows the contribution to the light yield from excitations and ionizations separately, assuming that each ionization produces one VUV photon in addition to successor excitation processes that contribute to the signal multiplication. One can observe that ionizations are relevant at very high reduced electric field and accordingly excitations increase exponentially as expected.    
The question on the non-linear behavior of the detector at E/p values lower than at those observed in other detectors remains unsolved. One of the near future plans is to modify our set-up to test for possible photoelectric effect in the gate mesh.

\subsection{Evaluation of the detection threshold}
\label{subsec:det_threshold}
Notwithstanding the exponential behavior, the observed \gtwo\ is sufficiently large to achieve a detection threshold extremely competitive. Concretely, the maximum charge yield is achieved at 8.62 bar and an electric field in the electroluminescent region of 1.52 kV/cm. A value of 711.6$\pm$58.6 photons per keV is observed with the uncertainty being fully dominated by the PMT gain decrease commented on section \ref{subsec:Fe}. This translates into a \gtwo\ value of 18.7$\pm$1.7 detected photons per electron, which should be more than sufficient to detect single electrons. 

However, the spectrum exhibits the presence of a low-energy population which results in an effectively higher threshold. For example, this population becomes dominant below $\sim$0.5 keV when operating at the maximum charge yield. The reconstructed energy for this population decreases with the electroluminescent yield, which points towards instrumental effects related to instabilities at high voltage (i.e. glows and/or field emission from the gate grid). 

Aside from this limitation, the optimal way to demonstrate the detection threshold would be by using calibration sources with as low energy as possible. Unfortunately, the lowest energy source available was \fe. Still, a reasonable assessment of the threshold can be given by evaluating the calibration peak at different yields and assuming that the number of photons detected at minimum yield are an effective threshold. Concretely, the rationale is that if the \fe\ peak can be observed at different electroluminescent yields, then the ratio between the yields can be used to estimate the minimum detection threshold at the higher yield. For example, given two datasets taken with a factor 10 difference in yield, the fact that the 5.9 keV can be observed in the lower yield dataset means that 0.59 keV events should be detected when operating at the higher yield. 

\begin{figure}
\centering
  \includegraphics{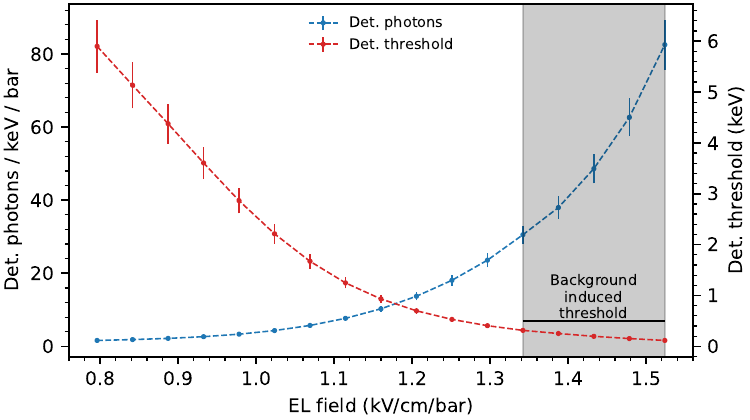} 
\caption{In blue, detected photons per keV and bar extracted from \fe\ peak at 8.62 bar. In red, the detection threshold assuming the minimum detectable signal corresponds to the number of detected photons per keV (14.00 $\pm$ 1.15 phot./keV) in the lower yield conditions. The shadowed grey area marks the point where the instrumental background dominates the threshold.
}
\label{fig:det_threshold} 
\end{figure}

The estimates for the extrapolated detection threshold are illustrated in Fig.~\ref{fig:det_threshold}, accompanied by the number of photons detected per keV and bar, for the data acquired at 8.62 bar. The lowest threshold corresponds to 0.117 $\pm$ 0.010 keV, around 4.5 ionization electrons, for an electroluminescence field of 1.52 kV/cm/bar. However, the low background population dominates the threshold down to a field of 1.34 kV/cm/bar. This population becomes subdominant at a field of 1.30 kV/cm/bar, where the detection threshold is estimated to be 0.42 $\pm$ 0.04 keV which corresponds to approximately 16 ionization electrons. Fig.~\ref{fig:fe_spectrum} shows an example of the threshold level in a spectrum. 

\subsection{Energy resolution}
\label{subsec:eres}

 The expected energy resolution of a gaseous detector with electroluminescence amplification was defined in \cite{dosSantos2001_Eres} and fundamentally depends on the intrinsic resolution of the gas (Fano factor) and the fluctuations in the amplification and photon detection process. Mathematically this results in equation \ref{eq:reso}:

\begin{equation}
    R_e = 2.35 \sqrt{\frac{F}{N_{e^-}} + \frac{\sigma_{EL}^2}{N_{e^-}\cdot Y_{EL}^2} + \frac{1 + (\sigma_{q}/q)^2}{N_{det}} + \frac{\sigma_b^2}{N_{det}^2}}% + \sigma_c^2}
    \label{eq:reso}
\end{equation}

In Eq.~\ref{eq:reso} each summand accounts for different fluctuations. Thus, the first describes fluctuations in the number of ionization electrons produced, with $F$ being the Fano factor, 0.23 $\pm$ 0.05 in gaseous Ar \cite{Ar_Fano}. The second term describes fluctuations in the electroluminescence process with $Y_{EL}$ being the absolute yield per electron and $\sigma_{EL}$ its variance. Its contribution is generally much smaller than the Fano factor and can be considered negligible as shown in \cite{OLIVEIRA2011217}. The third element defines fluctuations related to the sensor charge resolution. It is determined by the number of photons detected, $N_{det}$, and the sensor relative variance $(\sigma_{q}/q)^2$ being $q$ the average charge produced by a single photon and $\sigma_{q}$ its standard deviation. The fourth term is introduced to account for variations in the waveform baseline during the integration window, with $\sigma_b^2$ being the standard deviation anticipated based on the variation observed in each event's waveform, computed on a run-by-run basis.

\begin{figure}[tbp]
\centering
  \hspace{-0.1in}
  \includegraphics{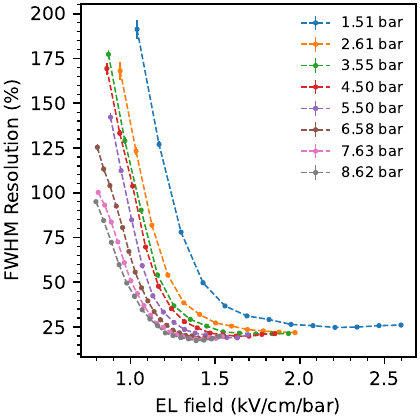}
  \hspace{0.1in}
  \includegraphics{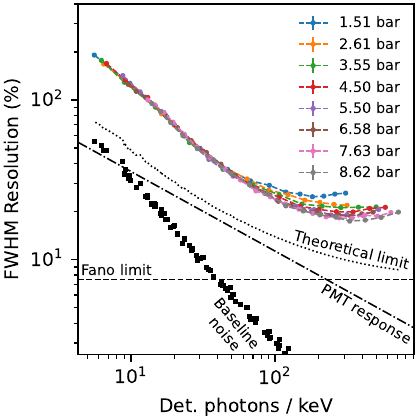}
\caption{Left: Dependence of the energy resolution with the reduced field in the EL region for different pressures. One can see that the resolution saturates for values of the reduced field in the region [1.25, 1.75] kV/cm/bar where the light produced is enough to minimize the statistical fluctuations. Right: Energy resolution as a function of the number of detected photons per keV for different pressures and its comparison with the theoretical limit calculated using \ref{eq:reso}. It can be observed that there is a clear difference between the observed and the expected resolution.}
\label{fig:reso} 
\end{figure}

In Fig. \ref{fig:reso} we compare the theoretical expected value for the resolution at the \fe\ peak as a function of the photons detected per keV. 
We can observe two effects. First, the resolution at low reduced electric fields evolves with a higher slope than the theoretical value, our main hypothesis for this is the fact that very few photons per electron are produced at these configurations as can be observed in Fig. \ref{fig:z_excitation} electrons can travel a few millimeters inside the EL gap before producing any excitation. In this situation the position of the few photons produced by the different electrons will be very different and, as the detector plane is close to the EL region, the collection efficiencies for different points of the EL are quite different. The combination of these two factors will account for an extra fluctuation parameter that disappears once the distance in between excitations inside the EL is reduced and therefore the light production is more homogeneous.
Second, at higher reduced fields, we observe a saturation of the energy resolution at a value that is about 80\% worse than the theoretical value.
We associate this difference with the non-linearity of the amplification region at large fields.
Both effects are planned to be investigated in the future with a combination of simulations and detector modifications.

\section{Conclusion and future plans}
\label{sec:conclusions}

This paper has described the Gaseous Prototype detector and its various subsystems. The detector has been first operated with gaseous argon to moderate pressures up to $\sim$10 bar. A 5.9 keV \fe\ source has been used to evaluate the detector's performance.

We observed an exponential behavior of the charge yield with the electroluminescence field which is not understood at the moment. At the same time, the energy resolution of the system appears to be $\sim$80\% worse than the theoretical value, a deviation we suspect to be related to the exponential yield. In an attempt to explain the non-linear behavior, a detailed simulation of the electric field in the electroluminescence was performed to identify possible intense field regions where charge amplification could occur. However, the simulation results do not deviate from the expected linear trend. We theorize that the photoelectric effect in the electroluminescence grid may be the origin of the charge amplification, a hypothesis we plan on evaluating in the near-future. 

In spite of these instrumental effects, the presented results mark the first demonstration of the low detection threshold that can be achieved with the HPNG EL-TPC technique, confirming its enormous promise for \cenuns\ detection. Concretely, a threshold as low as 0.117$\pm$0.010 $\rm{keV}_{\rm{ee}}$, has been estimated when operating at 8.62 bar, the maximum pressure considered. However, instrumental backgrounds, likely related to the high voltage operation, limit such threshold to 0.42$\pm$0.04 $\rm{keV}_{\rm{ee}}$. It should be noted that a more careful peak selection may filter out the background population and allow to recover the lower detection threshold. In any case, assuming a quenching factor in gaseous Ar similar to values reported for gaseous Xe \cite{renner_qf_xe}, the threshold would correspond to a nuclear recoil of 2.32 $\rm{keV}_{\rm{nr}}$ and, if the instrumental backgrounds can be suppressed, 0.65 $\rm{keV}_{\rm{nr}}$. These estimates will be further evaluated and corroborated in the future by using lower energy sources and measuring the quenching factor of gaseous Ar. Moreover, the studies will soon be extended to higher pressures as well as using Xe and ArXe mixtures.

\section*{Acknowledgments}

We are grateful with L. Arazi for some insightful discussions on this work. This project has received support from the European Research Council (ERC) under Grant Agreement No. 101039048-GanESS.
AS acknowledges support from the European Union’s Horizon 2020 research and innovation programme under the Marie Sk\l odowska-Curie grant agreement No 101026628. LL is supported by the predoctoral training program non-doctoral research personnel of the Department of Education of the Basque Government.

\bibliographystyle{IEEEtran}
\bibliography{biblio.bib}

\end{document}